# Generalized spiral transformation for high-resolution sorting of vortex modes

**Jie Cheng,[1] Chenhao Wan,[1,2,*] and Qiwen Zhan[2,*]**

[1] *School of Optical and Electronic Information and Wuhan National Laboratory for Optoelectronics, Huazhong University of Science and Technology, Wuhan, Hubei 430074, China*
[2] *School of Optical-Electrical and Computer Engineering, University of Shanghai for Science and Technology, Shanghai 200093, China*
*\* wanchenhao@hotmail.com, qwzhan@usst.edu.cn*

**Abstract:** We propose a generalized spiral transformation scheme that is versatile to incorporate various types of spirals such as the Archimedean spiral and the Fermat spiral. Taking advantage of the equidistant feature, we choose the Archimedean spiral mapping and demonstrate its application in high-resolution orbital angular momentum mode sorting. Given a fixed minimum spiral width, the Archimedean spiral mapping shows superior performance over the logarithmic spiral mapping. This generalized transformation scheme may also find various applications in optical transformation and can be easily extended to other fields related to conformal mapping.

## 1. Introduction

Vortex is a natural feature existing in a variety of physical phenomena spanning from spin of the Galaxy to rotation of molecules. It has been widely known that light can carry a spin angular momentum(SAM) of $\pm\hbar$ per photon[1, 2]. In 1992, Allen et al. pointed out that a vortex beam with a spiral phase structure in the form of $e^{il\varphi}$, can carry an orbital angular momentum(OAM) of $l\hbar$ per photon, where $\varphi$ is the azimuthal angle in the cross section and the integer $l$ is the topological charge[3]. Being theoretically infinite in the state space and mutually orthogonal, these OAM states of light provides a new dimension and boosts the capacity for optical communication in both classical[4-6] and quantum[7, 8] fields.

A number of methods have been demonstrated to generate light with different OAM states for optical communication, such as spiral phase plates[9, 10], forked grating holograms[11], q-plates[12], mode conversion[13, 14], and spin-orbit coupling [15]. On the receiver side, OAM mode sorter plays a critical role in OAM-based optical communication. One straightforward method is using a forked grating hologram[16]. As one hologram can measure only one specific state at a time, the efficiency of this method is restricted to 1/N, where N is number of the OAM states. The Mach-Zehnder (M-Z) interferometer OAM sorter is capable of sorting OAM modes at the level of a single photon. However, N different states requires a complex system of cascading N-1 interferometers. An OAM sorter based on the log-polar coordinate transformation circumvents such complexity gracefully and measures all OAM states simultaneously [17]. Unfortunately, the overlapping between adjacent states barely meets the Rayleigh criterion, deteriorating the signal-to-noise ratio (SNR) of neighboring channels. Incorporating an additional fan-out element can greatly decrease the overlapping and substantially increase the SNR [18]. The scheme can be integrated to a compact form for practical use [19]. One disadvantage of this additional diffractive optical element is that it reduces the overall efficiency of the system. To solve this problem in principle, Wen et al. has put forward a new coordinate transformation that maps logarithmic spirals to parallel lines[20]. The number of turns of the logarithmic spiral determines the multiples by which the resolution can be increased.

It is well known that the logarithmic spiral is an equiangular spiral, the outer spirals are increasingly and significantly wider than the inner spirals. Given an annular intensity distribution and a minimum spiral width, the number of turns of logarithmic spirals are quite limited as shown in Fig. 1(a). On the other hand, the Archimedes spiral, of equidistant feature, has equal width for every spiral turn. Therefore, given the same annular intensity distribution and the same minimum spiral width, the number of turns can be increased noticeably as shown in Fig. 1(b).

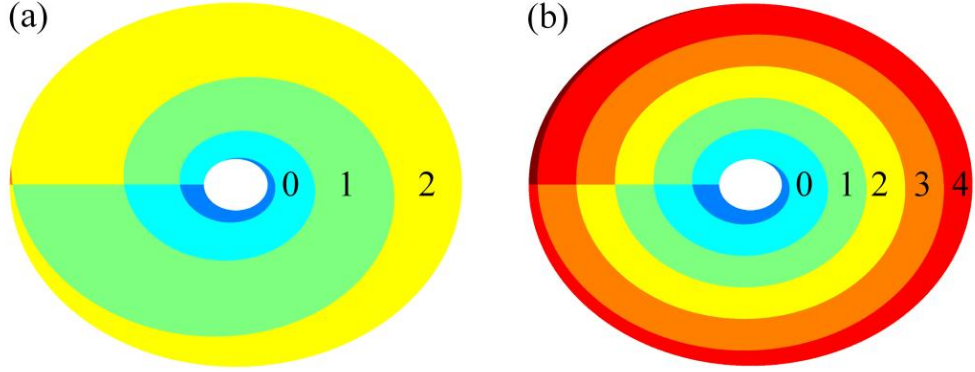

**FIG.1. Vortex beam decomposed by (a) logarithmic spiral and (b) Archimedes spiral. The number represents the turns of the spiral.**

In this work, we propose and demonstrate a general optical transformation scheme that can utilize various types of spiral mapping for sorting OAM states including the Archimedes spiral, the Fermat Spiral, etc. Taking advantage of the equidistant feature, we present detailed theoretical analysis, numerical simulation, and experimental demonstration for the Archimedes spiral OAM sorting. It is proven that the Archimedes spiral mapping can substantially increase the number of spiral turns for the mapping process and thus increase the resolution for OAM sorting. The general transformation scheme can also be extended to other research fields based on conformal mapping.

## 2. Theory

The optical mapping between two planes under paraxial approximation is expressed in terms of the phase gradient by the following equations[21]:

$$Q_x = k\frac{u-x}{d}, Q_y = k\frac{v-y}{d}, \quad (1)$$

where $(x, y)$ is the Cartesian coordinates of the input plane, $(u, v)$ is Cartesian coordinates of the output plane, $Q(x, y)$ is the phase distribution in the input plane, $Q_x$ and $Q_y$ are the partial derivatives with respect to the corresponding variables, $k$ is the free-space wave number, and $d$ is distance between the input plane and the output plane. New coordinates $(s, \theta)$ are introduced to replace the Cartesian coordinates $(x, y)$ in the input plane:

$$x = r(s,\theta)\cos(\theta), y = r(s,\theta)\sin(\theta), \theta = \varphi + 2m\pi, \quad (2)$$

where $(r, \varphi)$ are the polar coordinates of the input plane and $m$ is an integer indicating the number of spiral turns. The input plane is decomposed to a set of spirals of infinite length and finite width. A point located at $(x, y)$ can be represented by a parameter $s$ that indicates a particular spiral line that goes through the point and a parameter $\theta$ that indicates the unwrapped azimuthal angle along the spiral line. Since a spiral is of infinite length, the upper bound of the parameter $\theta$ goes to infinity. In other words, the range of angle extends from $[0, 2\pi)$ to $[0, +\infty)$.

The type of spirals is governed by the equation $r = r(s,\theta)$, e.g., an Archimedean spiral is expressed as $r = s + a_1\theta$, where $a_1 > 0$.

The underlying principle for OAM sorting based on spiral mapping is to convert the spiral phase of OAM states into a prism phase that can be later spatially separated by a simple lens. That is to say, the points with the same azimuthal angle $\varphi$ in the input plane must be mapped to the points with the same coordinate $u$ in the output plane:

$$u = u(\varphi), v = v(r,\varphi), \quad (3)$$

which can also be expressed by the spiral coordinates

$$u = u(\theta), v = v(s,\theta). \quad (4)$$

The coordinate $v$ depends both on the parameter $s$ and $\theta$.

To obtain one possible solution of the transformation phase function $Q(x,y)$, we assume that the phase function has a continuous second-order partial derivative, i.e., $Q_{xy} = Q_{yx}$. From Eq.(1), it is obtained that $u_y = v_x$. Combining with Eq.(4), the following equation can be derived,

$$v_s s_x + v_\theta \theta_x = u'(\theta)\theta_y. \quad (5)$$

Using the relationship between the Cartesian coordinates and the spiral coordinates (Eq.(2)), Eq.(5) can be rewritten as follows,

$$\left[\frac{v_s}{r_s}r - u'(\theta)\right]x + \left[\frac{v_s}{r_s}r_\theta - v_\theta\right]y = 0. \quad (6)$$

Eq. (6) is a condition that must be satisfied throughout the entire input plane. For simplicity, we set the expression in the parentheses equal to zero:

$$\begin{cases} u'(\theta) = r\dfrac{v_s}{r_s} \\ v_\theta = r_\theta \dfrac{v_s}{r_s} \end{cases}. \quad (7)$$

From Eq.(4), it is seen that the coordinate $u$ depends only on the angle $\theta$, which means that $\dfrac{du(\theta)}{d\theta} = f(\theta)$. Therefore, it is convenient to assume that $u'(\theta) = a$, i.e., $u = a\theta$, where $a$ is a constant that controls the size of the transformed beam. Using the above assumptions, the following relationship can be obtained,

$$\begin{cases} u'(\theta) = a \\ v_\theta = a\dfrac{r_\theta}{r} \\ v_s = a\dfrac{r_s}{r} \end{cases}. \quad (8)$$

For a spiral expressed in the form of $r = r(s,\theta)$, Eq.(8) can be integrated to yield

$$\begin{cases} u = a\theta \\ v = a\ln\dfrac{r}{b} \end{cases}, \tag{9}$$

where $\theta$ is the angle along the spiral line, $r$ is the distance between the point on the spiral line and the origin, and $b$ is the parameter that controls the longitudinal displacement of the transformed beam. Thus, the mapping condition that applies to various types of spirals is derived.

Through introducing the complex variables $Z = x+iy$ and $W = v+iu$, we can express the conformal mapping between the input plane (Z) and output plane (W) based on Eq.(9),

$$W = a(\ln\dfrac{Z}{b} + i2m\pi), \tag{10}$$

where $m$ is the number of spiral turns that is related to a specific spiral type. If $m$ is set to 0, the spiral coordinate will degenerate into an ordinary Cartesian coordinate, and the mapping will reduce to the traditional log-polar mapping.

The phase function can be obtained based on Eq.(1),

$$Q = \dfrac{ak}{d}\left[x\left(\arctan\dfrac{y}{x} + 2m\pi\right) + y\ln\dfrac{\sqrt{x^2+y^2}}{b} - y\right] - \dfrac{k(x^2+y^2)}{2d}, \tag{11}$$

where $k$ is the free-space wave number, $d$ is distance between input plane and output plane. For the Archimedean spiral, $m$ is expressed as,

$$m = \left\lfloor \dfrac{1}{2\pi}(\dfrac{r-r_0}{a_1} - \varphi) \right\rfloor, \tag{12}$$

where $\lfloor \ \rfloor$ stands for the integer part, $r_0$ is the radius of the first turn and $a_1$ is a constant controlling the spiral width of each turn.

In the output plane, another phase function should be applied to recollimate the beam. This phase function, generally known as the corrector phase, is expressed as [21],

$$P = \dfrac{abk}{d}e^{\frac{v}{a}}\sin\dfrac{u}{a} - \dfrac{k(u^2+v^2)}{2d}, \tag{13}$$

where $(u,v)$ are the Cartesian coordinates in the output plane. A lens is placed after the phase corrector to spatially separate light beams with different prism phase. The position of the focal spot depends on the gradient of the prism phase, or the topological charge $l$ of the original spiral phase,

$$u(l) = \dfrac{\lambda f l}{2\pi a}, \tag{14}$$

where $f$ is the focal length and $\lambda$ is the wavelength.

### 3. Numerical simulation and experimental results

A typical example of the transformation phase function and corrector phase function for the Archimedean spiral transformation are plotted in Fig.2(a) and Fig.2(b), respectively. It is obvious that there exists discontinuity in the transformation phase, which is caused by the introduction of the spiral coordinates. Consequently, there will be a minimum spiral width requirement in order to properly apply the optical transformation. The optical setup is shown in Fig.2(c). A He-Ne laser at 633nm is used as the light source. A polarizer is utilized to align the direction of linear polarization of the light with the horizontal direction because the SLM is responsive only to the horizontal direction. A spatial filter, composed of two lenses (L1 and L2) and a pinhole, filters out stray light from the laser. In this experiment, two SLMs are used. The first SLM is divided into two halves, the left half is used for generating the superposition of OAM states and the right half is used to display the transformation phase $Q(x,y)$. A small piece of black paper is placed between BS1 and BS2 to block the light from transmitting through. The second SLM is used to project the correction phase $P(u,v)$. Finally, the transformed beams with different prism phase gradient are focused by a lens to different horizontal positions on the CCD camera.

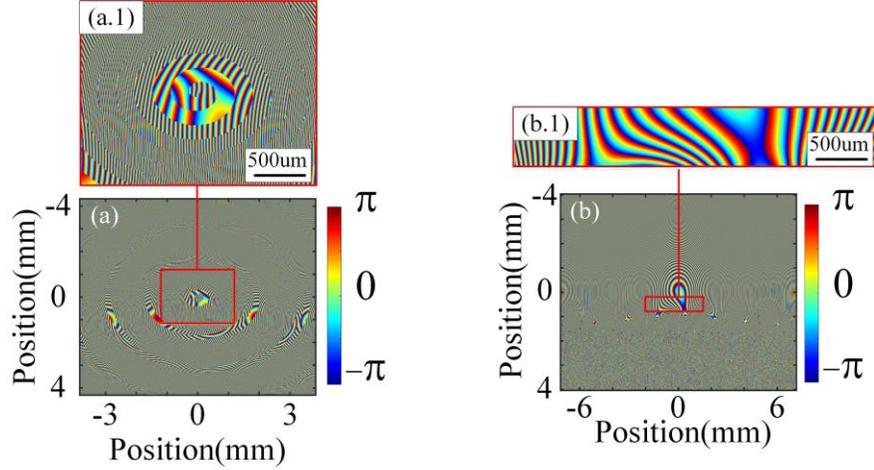

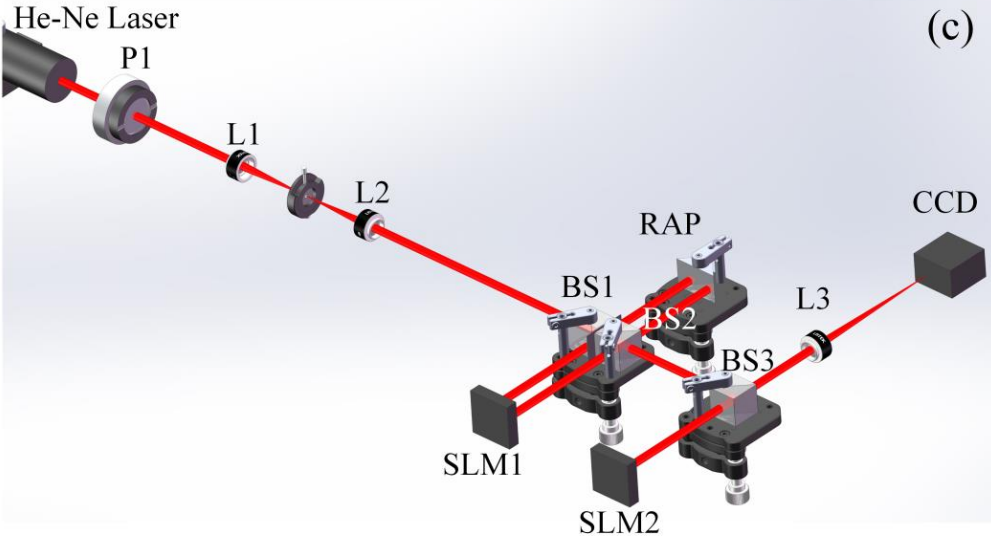

**FIG.2. Profile of (a) the transformation phase (Archimedean spiral) and (b) the corrector phase. (c) Schematic overview of the setup.** P1, Polarizer; L1, L2 and L3, Fourier lens; BS1, BS2 and BS3, Non-polarizing beam splitter; RAP, Right-angle prism; SLM1 and SLM2, Spatial light modulator. (a.1) and (b.1) are the details of the transformation phase and corrector phase respectively. The colorbar indicates the phase range in Fig.2(a) and Fig.2(b).

In this work, the minimum spiral width is setting equal to the radius of the first turn, that is $r_0$. The parameters being used are $d = 202.5$mm, $2\pi a = 2$mm, $r_0 = 0.183$mm, $a_1 = \dfrac{0.183}{2\pi}$ mm for the Archimedean spiral and $d = 202.5$mm, $2\pi a = 2$mm, $r_0 = 0.183$mm, $2\pi a_1 = \log(2)$ for the logarithmic spiral. The comparison between the logarithmic spiral transformation and the Archimedean transformation are shown in Fig. 3 and Fig. 4. The input light for numerical simulations is a perfect vortex beam with a fixed ring diameter [22].

Figure 3 shows the numerical and experimental results of the separated OAM modes with different topological charges ($-2 \leq l \leq 2$) using the logarithmic spiral transformation [Figs.3(a) and 3(b)] and the Archimedean spiral transformation (Fig.3(c) and 3(d)). Figure 4 depicts the intensity linescans of separated modes along the red dashed lines in Fig.3. Different OAM modes are separated by a distance that is proportional to the topological charge $l$.

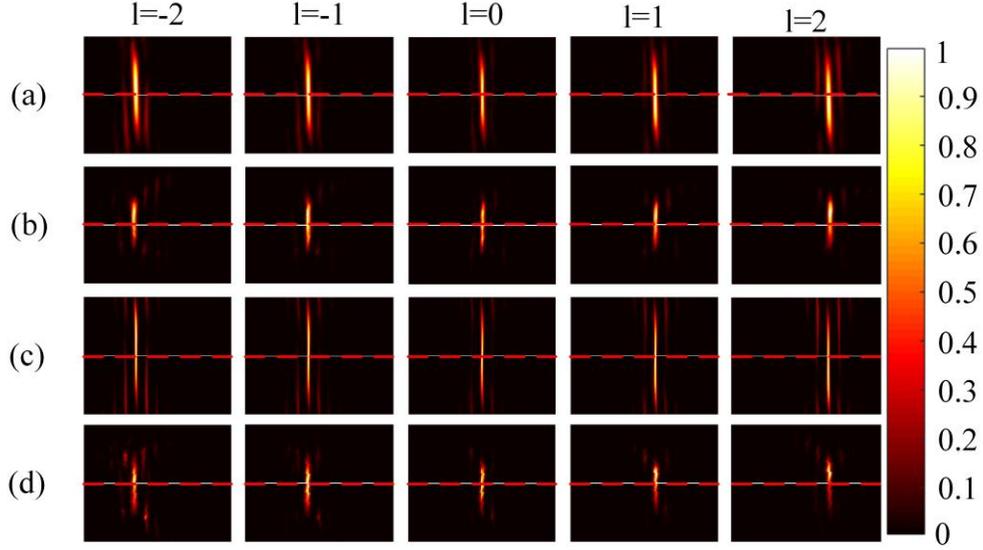

**FIG.3. Separated OAM modes with different topological charges ($-2 \leq l \leq 2$).** (a) numerical simulations and (b) experiments for the logarithmic spiral transformation. (c) numerical simulations and (d) experiments for the Archimedean spiral transformation. The colorbar indicates the intensity of different OAM modes. All images have the same spatial scale.

The optical finesse F defined as the ratio of the spacing between adjacent OAM states over their average FWHM [19], is used to quantify the resolution for sorting different OAM modes. Based on the simulation results, the F is 2.987 for the logarithmic spiral transformation and is 6.190 for the Archimedean spiral transformation (Fig. 4(a) and 4(c)). That is to say, given the same annular intensity distribution and minimum spiral width, the Archimedean spiral mapping shows an OAM-sorting-resolution twice better than that of the logarithmic spiral mapping. The increase in resolution is proportional to the number of turns of spirals in the input plane. It is expected that given a larger beam size and the same minimum spiral width, the Archimedean spiral mapping will perform increasingly better than the logarithmic spiral mapping. Based on the experimental results, the F is 2.332 for the logarithmic spiral transformation and is 2.891 for the Archimedean spiral transformation (Fig. 4(b) and 4(d)). The resolution increase is limited to only 24% which is caused by the limited width of the SLM. The transformed beam of the Archimedean spiral is approximately twice the length of the transformed logarithmic spiral, and is longer than the width of the SLM being used. Therefore, the experimental results actually show the resolution for a truncated transformed Archimedean spiral and the sorting resolution is not as good as the numerical simulation. The transformed beam for the Archimedes spiral mapping has a better intensity uniformity, resulting in a better focal spot. The resolution improvement of the Archimedes spiral mapping over the logarithmic spiral mapping can be calculated analytically. Setting the minimum spiral width to $\alpha r_0$, the width of the circular incident light to $\beta r_0$, the Archimedean spiral mapping will have a finesse $\dfrac{\beta/\alpha}{\log_2(\beta/\alpha)}$ times better than that of the logarithmic spiral mapping, where $\dfrac{\beta/\alpha}{\log_2(\beta/\alpha)}$ is always greater than 1.

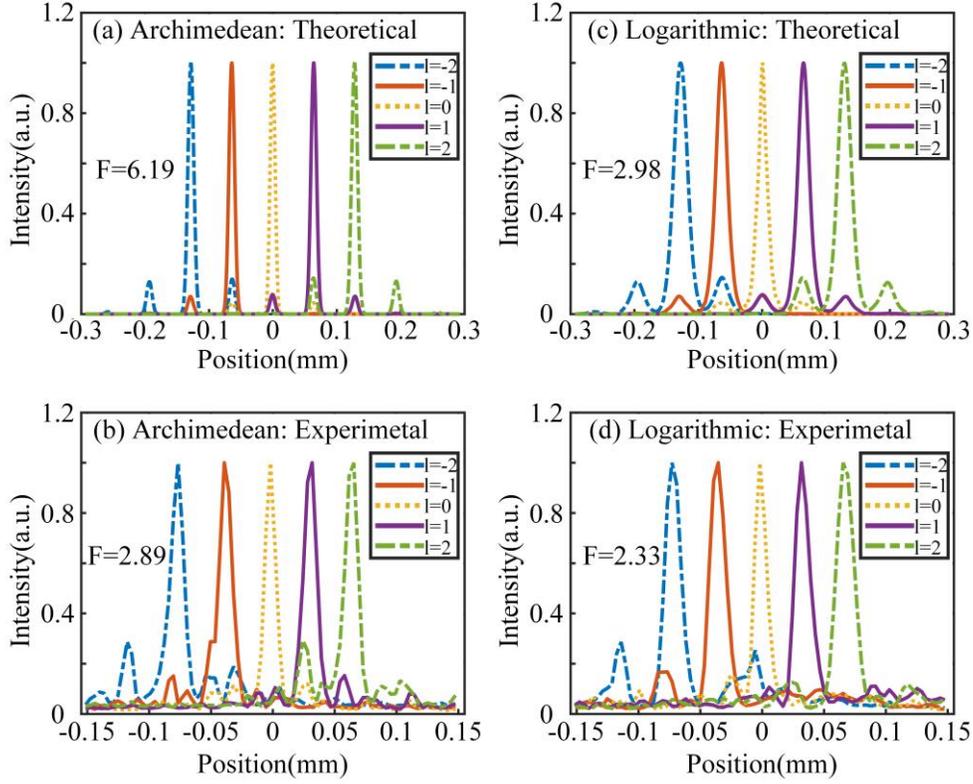

**FIG.4. Intensity distribution along the red dashed lines in the images of Fig.3 for different OAM modes.** (a) numerical simulations and (b) experiments for the Archimedean spiral transformation. (c) numerical simulations and (d) experiments for the logarithmic spiral transformation.

### 4. Discussions and conclusions

Despite the significant improvement in sorting different OAM states, it is found from Fig. 4 that the intensity of the final spot is not a strict $\sin c^2(x)$ function, which is mainly caused by the inherent properties of the vortex beam. Since the transformation phase is calculated under the premise of normally incident beam, using vortex beam as incident beam will introduce a skew angle being $l/kr$, which will result in a sinusoidal distortion on the transformed beam[17]. One method to reduce this distortion is by increasing the phase gradient of the transformation phase. By doing this, the angle introduced by the transformation phase will become much more significant than the skew angle introduced by the vortex beam. Also, decreasing the distance between the two phase elements can reduce this distortion as well.

It is recognized that the beam splitters used in the system cause a significant energy loss. This energy loss can be alleviated by replacing the SLMs with diffractive optical elements (DOEs) and changing from reflective to transmissive system. Through integrating the two phase elements into one component, the intensity distortion caused by misalignment can be reduced and the system can be made more compact and integratable[23, 24].

Except for the Archimedean spiral, the new transformation scheme is flexible in exploiting different types of spirals, such as the Fermat spiral, which is determined by the formula $r = s + a_1\sqrt{\theta}$. For optical vortex communication system with Laguerre-Gaussian beam source, the ring radius of the incident beam is proportional to the topological charge $l$. Therefore, beneficial from the narrower and narrower spiral width from inner to outer shells of the Fermat spiral, this spiral mapping will find its unique advantage in this situation. However, narrow spiral width also means a high requirement for SLM resolution, therefore, the Fermat spiral mapping doesn't perform very well in our system. This problem can be solved by using high-resolution DOEs or metamaterials.

In conclusion, we have proposed and demonstrated a generalized optical transformation scheme that is suitable for various types of spiral mapping. In particular, we choose the Archimedean spiral mapping due to its equidistant feature and demonstrate its application in high-resolution OAM mode sorting. Both the theoretical and the experimental results have shown a better performance of the Archimedean spiral mapping over the logarithmic spiral mapping. The generalized transformation method proposed in this paper lays a backbone for applications in various disciplines related to conformal mapping.

## Funding
This work was supported by the National Natural Science Foundation of China (92050202, 61875245), and Wuhan Science and Technology Bureau (2020010601012169).

## Author contribution
J.C., C.W, and Q.Z. proposed the original idea and performed the theoretical analysis. J.C. performed the experiments. Q.Z. guided the theoretical analysis and supervised the project. All authors contributed to writing the manuscript.

## Competing financial interest
The authors declare no competing financial interests.

## Data availability
All data of this study are available from corresponding authors upon reasonable request.

## Code availability
All codes used for data analysis and simulations are available from corresponding authors upon reasonable request.